\documentclass[12pt]{iopart}
\usepackage{epsf}

\newcommand{\simgt}{\lower.5ex\hbox{$\; \buildrel > \over \sim \;$}}
\newcommand{\simlt}{\lower.5ex\hbox{$\; \buildrel < \over \sim \;$}}

\begin{document}

\title[]{The finite source size effect and the wave optics in gravitational lensing}


\author{Norihito Matsunaga and Kazuhiro Yamamoto}
\address{
Graduate School of Sciences, Hiroshima University, 
Higashi-hiroshima, 739-8526, Japan}


\begin{abstract}
We investigate the finite source size effect in the context 
of the wave optics in the gravitational lensing. 
The magnification of an extended source is presented 
in an analytic manner for the singular isothermal sphere lens model 
as well as the point mass lens model with the use of the 
thin lens approximation. 
The condition that the finite source size effect 
becomes substantial is demonstrated. As an application, 
we discuss possible observational consequences of the 
finite source size effect on astrophysical systems.
\end{abstract}
\section{Introduction}

Gravitational lensing is a characteristic phenomenon of the
general relativity and has become a very important tool 
in the fields of cosmology and astrophysics \cite{SEF,FT}. 
For example, the existence of the massive compact halo 
objects (MACHO) in the Galaxy was revealed by the detection of 
the lensed amplification of stellar objects \cite{Alcock}, and the recent 
measurements of the cosmic shear field provide a constraint 
on the matter distribution independently of the 
clustering bias \cite{Bacon,Wittman,Kaiser,Maoli}. 
Promisingly the gravitational lensing will play a more important role 
with progress in the capability of observational facilities in future. 
For example, it will be a useful probe of the nature of the dark energy 
\cite{YF,YKMF,LSST,Bartelmann}.

As a fundamental aspect of the gravitational lensing, 
the effect of wave optics has been investigated by many authors 
\cite{Peters:1974gj,DeguchiA,DeguchiB,SEF}. 
Very recently, this subject is revisited by several authors, 
motivated by a possible phenomenon which might be
observed in the future gravitational wave experiments 
\cite{DN,TTN,BHN,PINQ,Ruffa,TN,Takahashi,TNB,TSM,JPM,KYA,KYB,KYC}.
In the context of the wave optics of gravitational lensing, the 
argument on the 
distance-redshift relation is also revisited \cite{Yoo}. 

The present paper focuses on the finite source size effect in the
wave optics of the gravitational lensing. In the first half part
of this paper, we present an analytic solution for the wave 
equation with the singular isothermal sphere lens model. 
In general, it is difficult to obtain an exact solution for general 
lens model, excepting a few simple lens models \cite{Suyama,KYB}. 
Therefore it is useful to obtain such analytic solution for the wave 
equation. The wave optics of the singular isothermal 
sphere lens has been investigated by Takahashi and Nakamura using a 
numerical technique \cite{Thesis,TNSIS}. 
We present the analytic expression for the amplification factor, 
which is the first aim of the present paper. 
Then, as an application of the analytic formula, we 
investigate the finite source size effect
in the wave optics, which is the other aim of the present paper. 
Using the analytic formula, we consider  the energy spectrum 
from an extended source with a 
Gaussian distribution of surface brightness \cite{Krzysztof}. 
We investigate the condition that the finite source 
size effect becomes important in the wave optics, 
including the case near the caustic in the limit of the 
geometrical optics. 

This paper is organized as follows: In section 2, we review the 
basic formulas for the wave optics in gravitational lensing. 
The limit of the geometrical optics is also reviewed for 
self-containment. 
Then, we present analytic expressions of the amplification 
factor for the point mass lens model as well as the singular 
isothermal sphere lens model in section 3. 
In section 4, we investigate the finite source size effect
with the use of the analytic formulas.
In section 5, the validity of the approximation of the point 
source is discussed for the gravitational wave from a compact
binary. The femtolensing of the gamma ray burst source is also 
revisited \cite{Krzysztof,Gould}, and the finite source size 
effect is considered.
The last section is devoted to summary and conclusions. 
Throughout this paper, we use the unit in which 
the light velocity equals $1$.

\section{Review of Basic Formalism}
\subsection{Wave Optics under the Thin Lens Approximation}

We consider the background spacetime with the line element,
\begin{eqnarray}
ds^{2} =g_{\mu \nu} dx^{\mu} dx^{\nu}=-(1+2U({\vec{r}}))dt^2
+(1-2U({\vec{r}}))d {\vec{r}}^2,
\label{MNEQN1}
\end{eqnarray}
where $U(\vec{r})$ is the Newtonian gravitational potential with
the condition $U(\vec{r})\ll1$.  
On the Newtonian background spacetime, we consider the 
wave propagation of the scalar field $\Phi$. The propagation 
of the electro-magnetic wave and the gravitational wave can 
be well described by the scalar wave equation \cite{Peters:1974gj,SEF},
which is given by
\begin{eqnarray}
\partial_\mu( \sqrt{-g} g^{\mu \nu} \partial_\nu \Phi)=0.
\end{eqnarray}
This is rewritten as
\begin{eqnarray}
(\nabla^{2}+{\omega}^{2})\Phi=4{\omega}^{2}U({\vec{r}})\Phi, 
\label{MNEQN2}
\end{eqnarray}
on the spacetime with the line element (\ref{MNEQN1}), where
we assume the monochromatic wave with the angular frequency $\omega$.
In the present paper we consider the spherically symmetric potential.

\begin{figure}[t]
\epsfxsize=10cm
\epsfbox{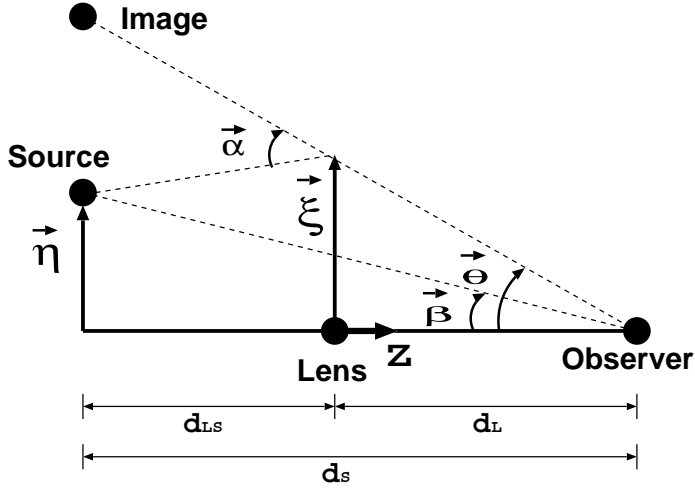}
\caption{\label{fig:fthin} 
Configuration of the source, the lens and the
observer. $d_{L},d_{S}$ and $d_{LS}$ are the distances between the 
lens and the observer, the source and observer, and the lens and the 
source, respectively.
$\vec{\eta}$ is the position of the (point) source, and 
$\vec{\xi}$ is the impact parameter. 
$\vec{\beta}$ is the unlensed source position angle, 
$\vec{\theta}$ is the position angle of the image, 
and the deflection angle is $\vec{\alpha}$. 
This sketch is based on the 
thin lens approximation that the wave is 
scattered only on the thin lens plane.} 
\label{fthin} 
\vspace{0.2cm}
\end{figure}

It is useful to introduce the amplification factor $F$ (which is called 
the transmission factor in Ref. \cite{SEF}) by $F=\Phi/{\Phi}_{0}$, 
where $\Phi_0$ is the wave amplitude in the absence of the 
gravitational potential, $U=0$.
Then, under the thin lens approximation, the amplification factor
is given by 
\begin{eqnarray}
F(\omega,\vec{\eta})=\frac{d_{S}}{d_{L}d_{LS}}\frac{\omega}{2\pi i} 
\int_{-\infty}^{\infty} 
d^2 \xi \exp[i\omega \hat{\phi}(\vec{\xi},\vec{\eta})],
\label{MNEQN3}
\end{eqnarray}
where $\hat{\phi}(\vec{\xi},\vec{\eta})$ is the time delay 
function (Fermat's potential), which is given by
\begin{eqnarray}
\hat{\phi}(\vec{\xi},\vec{\eta})=\frac{d_{L}d_{S}}{2d_{LS}}
\Biggl(\frac{\vec{\xi}}{d_{L}}-\frac{\vec{\eta}}{d_{S}}\Biggr)^{2}
-\hat{\psi}(\vec{\xi})
,\label{MNEQN4}
\end{eqnarray}
where $d_{L}$ is the distance between the lens and the source, 
$d_{S}$ is the distance between the source and the observer, 
$d_{LS}$ is the distance between the lens and the source, 
respectively,  (see Figure \ref{fig:fthin} 
for the configuration of the lensing system and the definition of variables). 
In general, we may add a term $\hat{\phi}_{m}(\vec{\eta})$ in the right 
hand side of (\ref{MNEQN4}) \cite{Thesis}.
However, the inclusion does not alter our arguments and we omit it. 
The two dimensional gravitational deflection potential 
is defined by
\begin{eqnarray}
\hat{\psi}({\vec{\xi}})=2 \int_{-\infty}^{\infty}dz 
U(\vec{\xi},z).\label{MNEQN5}
\end{eqnarray}
Note that $|F|=1$ in the absence of the lens potential $U=0$.

The above formulas can be generalized so as to take the cosmological 
expansion into 
account. Assuming that the wavelength of the scalar waves is much shorter 
than the horizon scale, Eq.~(\ref{MNEQN3}) is generalized as
\begin{eqnarray}
F(\omega,\vec{\eta})=\frac{d_{S}}{d_{L}d_{LS}}\frac{\omega(1+z_{L})}
{2\pi i}
\int_{-\infty}^{\infty}d^{2}\xi \exp[i\omega(1+z_{L})
\hat{\phi}(\vec{\xi},\vec{\eta})],\label{MNEQN6}
\end{eqnarray}
where $d_{L},~d_{S}$, and $d_{LS}$ are the angular diameter distances,
and $z_{L}$ is the redshift of the lens object. 

It is useful to rewrite the amplification factor $F$ in terms of 
dimensionless quantities. We introduce
\begin{eqnarray}
{\vec{x}}&=&\frac{\vec{\xi}}{\xi_{0}},\hspace{3mm}{\vec{y}}
=\frac{d_{L}}{\xi_{0}d_{S}}\vec{\eta},\label{MNEQN7} 
\\
w&=&\frac{d_{S}}{d_{L}d_{LS}}\xi_{0}^{2}(1+z_{L})\omega,\label{MNEQN8} \\
\psi&=&\frac{d_{L}d_{LS}}{d_{S}\xi_{0}^{2}}\hat{\psi},\label{MNEQN9}
\end{eqnarray}
where $\xi_{0}$ is the normalization constant of the length in the lens 
plane, 
for which we adopt 
\begin{eqnarray}
\xi_{0}=\theta_{E}d_{L},
\end{eqnarray}
where $\theta_{E}$ is the 
Einstein angle, i.e., the solution of the lens equation
(\ref{MNEQNthirteen})
with $\vec{\beta}=0$. (see below.)
The effect of the wave optics is characterized by the dimensionless 
parameter $w$.
We also introduce the dimensionless time delay function by 
\begin{eqnarray}
T({\vec{x}},{\vec{y}})=\frac{d_{L}d_{LS}}{d_{S}\xi_{0}^{2}}
\hat{\phi}(\vec{\xi},\vec{\eta})
=\frac{1}{2}|{\vec{x}}-{\vec{y}}|^{2}-\psi({\vec{x}}).\label{MNEQN10}
\end{eqnarray}
Then, the amplification factor is written as
\begin{eqnarray}
F(w,{\vec{y}})=\frac{w}{2\pi i}\int_{-\infty}^{\infty}d^{2}x
\exp[iwT({\vec{x}},{\vec{y}})].\label{MNEQN11}
\end{eqnarray}

In the case of the spherically symmetric lens model,
the gravitational deflection potential $\psi({\vec{x}})$ 
depends only on $x=|{\vec{x}}|$. Then, 
the amplification factor is reduced to the relatively simple formula
\begin{eqnarray}
F(w,y)=-iw e^{\frac{i}{2}w{y^2}}\int_{0}^{\infty}dx\hspace{1mm}x\hspace{1mm}
J_{0}(wxy)\exp\Biggl[iw\Biggl(\frac{1}{2}x^{2}-\psi(x)\Biggr)\Biggr],
\label{MNEQN12}
\end{eqnarray}
where $J_{0}(z)$ is the Bessel function of the zeroth order and 
$y=|\vec{y}|$.

\subsection{Geometrical Optics Approximation}
In this subsection we consider the limit of the short wave 
length in the wave optics $(w \gg 1)$,  
which reproduces the conventional geometrical 
optics in the gravitational lensing.
In the limit of the geometrical optics, the diffraction integral 
(\ref{MNEQN11}) is evaluated around the stationary points of the time 
delay function $T({\vec{x}},{\vec{y}})$. Thus the stationary points are 
determined by the solution of 
$\nabla_{x}T({\vec{x}},{\vec{y}})=0$, which is written as
\begin{eqnarray}
{\vec{y}}={\vec{x}}-\nabla_{x}\psi({\vec{x}}).
\label{MNEQN13}
\end{eqnarray}
This is the lens equation to determine the image position ${\vec{x}}_{j}$.
Eq.~(\ref{MNEQN13}) is rewritten as
\begin{eqnarray}
\vec{\beta}=\vec{\theta}-\vec{\alpha}(\vec{\theta}),
\label{MNEQNthirteen}
\end{eqnarray}
where $\vec{\beta}=(\xi_{0}/d_{L})\vec{y}$ is the angular position 
of the source,
$\vec{\theta}=(\xi_{0}/d_{L})\vec{x}$ is the angular position of the 
images, 
and $\vec{\alpha}=(\xi_{0}/d_{L})\nabla_{x}\psi(\vec{x})$ is the deflection 
angle (see Figure \ref{fig:fthin}). 

The time delay function $T({\vec{x}},{\vec{y}})$ is expressed 
around the j-th image position ${\vec{x}}_{j}$ as
\begin{eqnarray}
T({\vec{x}},{\vec{y}})&=&T({\vec{x}}_{j},{\vec{y}})
+\frac{1}{2}\sum_{a,b=1,2}\partial_{a}\partial_{b}
T({\vec{x}}_{j},{\vec{y}}){X}_{a}{X}_{b}+\mathcal{O}({X}^{3}),
\label{MNEQN14}
\end{eqnarray}
where $\vec{X}={\vec{x}}-{\vec{x}}_{j}$. Here, the term in proportion to 
$\vec{X}$ vanishes because $\vec{{x}}_{j}$ is the stationary point of 
$T({\vec{x}},{\vec{y}})$.
Inserting Eq.~(\ref{MNEQN14}) into Eq.~(\ref{MNEQN11}), 
we obtain the amplification factor in geometrical optics limit
\cite{DN,Thesis,TNSIS}
\begin{eqnarray}
F_{geo}(w,{\vec{y}})=\sum_{j}|\mu(\vec{x}_{j})|^{1/2}
\exp\biggl[iwT({\vec{x}}_{j},{\vec{y}})-i\frac{n_{j}}{2} \pi \biggr],
\label{MNEQN15}
\end{eqnarray}
where the magnification of the j-th image is 
$\mu(\vec{x}_{j})=1/\det(\partial{\vec{y}}/\partial{\vec{x}}_{j})$
and $n_{j}=0,1,2$ when ${\vec{x}}_{j}$ is a minimum, saddle, 
maximum point of $T({\vec{x}},{\vec{y}})$, respectively.
For the case of multi-lensed images, in the geometrical optics 
approximation,
the expression (\ref{MNEQN15}) means that the observed wave is described by 
a superposition of each wave with the amplitude, 
$|\mu(\vec{x}_{j})|^{1/2}$,
and the phase, $wT(\vec{x}_{j},\vec{y})-(n_{j}\pi)/2$ \cite{DN}.

\section{Analytic Expressions for the Amplification Factor}
In this section we present analytic expressions for the amplification factor 
(\ref{MNEQN12}) for the point mass lens model and the singular isothermal 
sphere (SIS) lens model. 
The expression for the point mass lens model is well known 
\cite{Peters:1974gj,DeguchiA,DeguchiB,SEF}. Recently the amplification 
factor for the SIS lens model is investigated by Takahashi and Nakamura 
with the use of a numerical method \cite{Thesis,TNSIS}. 
However, we derive an analytic expression for the SIS 
lens model in the present paper. The SIS lens model is often used to study 
a lensing phenomenon by a galaxy halo and by a cluster of galaxies. In order 
to understand the wave effect in the lensing by a halo, such an analytic 
formula is useful. The amplification factor obtained using the analytic
expression for the SIS lens model gives us the results consistent 
with those by Takahashi and Nakamura through their 
numerical method \cite{Thesis,TNSIS}. 

\subsection{Gravitational Deflection Potential}
Let us here summarize the relation between the gravitational 
deflection potential $\psi({\vec{x}})$ and the density distribution 
of a lens object (see e.g., \cite{SEF}). 
In general the deflection potential is given by 
\begin{eqnarray}
\psi({\vec{x}})=4G\frac{d_{L}d_{LS}}{d_{S}}\int_{-\infty}^{\infty}
d^{2}{s}\hspace{1mm}\Sigma({\vec{s}})\log|{\vec{x}}-{\vec{s}}|,
\label{MNEQN16}
\end{eqnarray}
where $\Sigma(\vec{s})$ is surface mass density in lens plane,
\begin{eqnarray}
\Sigma(\vec{x})=\int_{-\infty}^{\infty}dz \rho(\vec{x},z),
\end{eqnarray}
where $\rho(\vec{x},z)$ is the mass density distribution of the lens,
which is related to the Newtonian potential by $\nabla^{2}U=4\pi G\rho$. 
Thus the lens model is characterized by the mass density distribution 
$\rho(\vec{x},z)$ as well as the gravitational 
deflection potential $\psi({\vec{x}})$.  

\subsection{Point Mass Lens}
The gravitational lensing by a black hole and a compact star is described 
by the point mass lens model, in which we write
$\rho(\vec{x},z)=M\delta^{(2)}(\vec{\xi})\delta^{(1)}(z)$, 
where $M$ is the mass of the lens object.  Then, the surface mass density is 
$\Sigma({{\vec{x}}})
=M\delta^{(2)}({{\vec{\xi}}})=M\delta^{(2)}(\xi_{0}{{\vec{x}}})$. 
In this model the characteristic Einstein angle $\theta_{E}$ is 
\begin{eqnarray} 
\theta_{E}=\sqrt{\frac{4GM d_{LS}}{d_{L}d_{S}}}
\simeq 3\times10^{-6}\biggl(\frac{M}{M_{\odot}}\biggr)^{1/2}
\biggl(\frac{d_{L}d_{S}/d_{LS}}{\rm 1Gpc}\biggr)^{-1/2}
\hspace{2mm}{\rm arcsec},
\label{MNEQN18}
\end{eqnarray}
and the gravitational deflection potential 
is given by $\psi(x)=\log{x}$.
Using a mathematical integral formula \cite{Abramowitz},
the expression of the amplification factor (\ref{MNEQN12}) yields
\begin{eqnarray}
F=e^{\frac{i}{2}w(y^2+\log(w/2))}e^{\frac{\pi}{4}w}
\Gamma\biggl(1-\frac{i}{2}w\biggr)
{}_{1}F_{1}\biggl(1-\frac{i}{2}w,\hspace{1mm}1;\hspace{1mm}
-\frac{i}{2}wy^{2}\biggr),\label{MNEQN17}
\end{eqnarray}
where ${}_{1}F_{1}(a,c,z)$ is the confluent hypergeometric function
\cite{Magnus}. In this model we have the dimensionless parameter
from Eq.~(\ref{MNEQN8}), which characterizes the wave optics,
\begin{eqnarray}
w=4GM(1+z_{L})\omega \simeq 1.2
\times10^{-4}(1+z_{L})\biggl(\frac{M}{M_{\odot}}\biggr)
\biggl(\frac{\nu}{\rm 1Hz}\biggr).
\label{MNEQN19}
\end{eqnarray}
Note that $w$ has the meaning of the ratio of the Schwarzschild radius 
to the wavelength of the propagating wave. The wave effect becomes 
significant when $w\sim {\cal O}(1)$. 

\begin{figure}[t]
\epsfbox{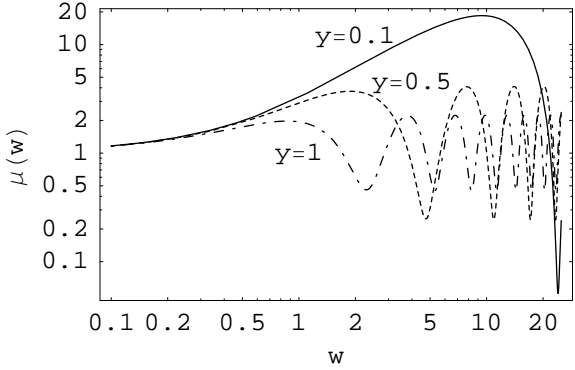}
\caption{\label{fig:two}
Magnification $\mu(w,y)$ as a function of dimensionless parameter 
$w$ for the point mass lens model. Here, the source 
position is fixed as $y=0.1,~0.5$, and $1$, respectively. 
}
\vspace{0.2cm}
\end{figure}

\begin{figure}
\epsfbox{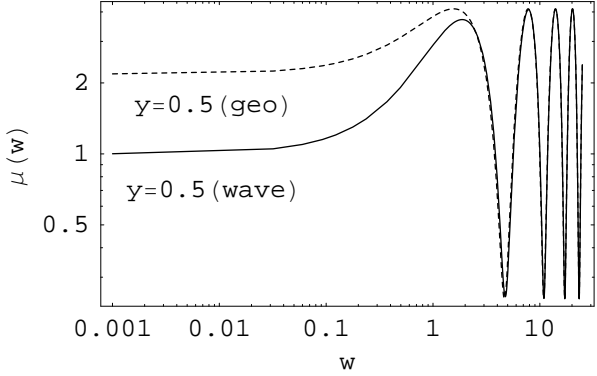}
\caption{\label{fig:three}
Magnification as a function of the dimensionless parameter $w$. 
The solid curve is the result of the wave optics $\mu(w,y)$,
while the dashed curve is $\mu_{geo}(w,y)$. 
Here, the source position is fixed as $y=0.5$.}
\vspace{0.2cm}
\end{figure}

We define the magnification by $\mu(w,y)\equiv|F(w,y)|^{2}$, which 
gives us the expression 
\begin{eqnarray}
\mu(w,y)=\frac{\pi w}{1-e^{-\pi w}}\Big|{}_{1}F_{1}\biggl(\frac{i}{2} 
w,\hspace{1mm}1;\frac{i}{2}wy^{2}\biggr)\Big|^{2},\label{MNEQN20}
\end{eqnarray}
from expression (\ref{MNEQN17}). The maximum magnification 
is achieved when $y=0$, which provides the configuration of 
the Einstein ring, 
\begin{eqnarray}
\mu_{max}=\frac{\pi w}{1-e^{-\pi w}}.
\label{21}
\end{eqnarray}
Figure \ref{fig:two} shows the magnification (\ref{MNEQN20}) 
as a function of the wave characteristic parameter $w$ with the source 
position fixed $y=0.1,~0.5$, and $1$. For $w\simgt 1$, the oscillation
feature appears due to the interference in the wave effect
between the double images (see also Figure \ref{fig:three}).

We next consider the approximation based on the geometrical optics 
explained in the previous section. 
The point mass lens model has the two images in the geometrical optics.  
Namely, the lens equation has the two solution
(the minimum and the saddle points of the time delay function).
Then, (\ref{MNEQN15}) yields  
\begin{eqnarray}
F_{geo}(w,\vec{y})&=&
|\mu_{+}|^{1/2}\exp\biggl[i w\biggl(\frac{1}{2}(p_{+}-y)^{2}
-\log|p_{+}|\biggr)\biggr]
\nonumber\\
&-i& |\mu_{-}|^{1/2}\exp\biggl[i w\biggl(\frac{1}{2}(p_{-}-y)^{2}
-\log|p_{-}|\biggr)\biggr],\label{22}
\end{eqnarray}
where the magnification of each image is
$\mu_{\pm}=1/2 \pm (y^{2}+2)/(2y\sqrt{y^{2}+4})$ 
and $p_{\pm}=(1/2)(y \pm \sqrt{y^{2}+4})$. 
Then, the corresponding magnification is 
\begin{eqnarray}
\mu_{geo}(w,y)
&=&\frac{y^{2}+2}{y\sqrt{y^{2}+4}}
\nonumber \\ 
&+&\frac{2}{y\sqrt{y^{2}+4}}
\sin\biggl[w\biggl(\frac{1}{2}y\sqrt{y^{2}+4}
+\log\Big|\frac{\sqrt{y^{2}+4}+y}{\sqrt{y^{2}+4}-y}\Big|\biggr)\biggr].
\label{MNEQN23}
\end{eqnarray}
Figure \ref{fig:three} shows the magnification (\ref{MNEQN20}) 
and (\ref{MNEQN23}), 
as a function of the parameter $w$ with the source position fixed $y=0.5$.
For $w\simgt 1$, both the curves agree, and the geometrical optics 
is a very good approximation. For $w\simlt 1$, however, the two 
curves are not in good agreement because the geometrical optics 
approximation is not suitable. 

\subsection{Singular Isothermal Sphere Lens}
We next consider the SIS lens model, which can be used for 
modeling a halo. In this model, the density profile is 
\begin{eqnarray}
\rho(\vec{x},z)={\sigma_{v}^{2}\over 2\pi G(|\vec{\xi}|^{2}+z^{2})}
={\sigma_{v}^{2}\over 2\pi G(|\xi_{0}\vec{x}|^{2}+z^{2})},
\end{eqnarray}
where $\sigma_{v}$ is the velocity dispersion. 
Then, the surface density is given by
\begin{eqnarray}
\Sigma({\vec{x}})
=\frac{\sigma_{v}^{2}}{2G|\vec{\xi}|}
=\frac{\sigma_{v}^{2}}{2G\xi_{0}x}, 
\end{eqnarray}
and the gravitational deflection potential is given by 
$\psi(x)=x$. 
The Einstein angle of the SIS lens model is 
\begin{eqnarray}
\theta_{E}=4\pi \sigma_{v}^{2}\frac{d_{LS}}{d_{S}}
\simeq 3\times10^{-5}\biggl(\frac{\sigma_{v}}{\rm 1km/s}\biggr)^{2}
\biggl(\frac{d_{LS}}{d_{S}}\biggr)\hspace{2mm}{\rm arcsec},\label{MNEQN25}
\end{eqnarray}
therefore, the dimensionless parameter $w$ is given by
\begin{eqnarray}
w&=&(1+z_{L})\omega(4\pi \sigma_{v}^{2})^{2}\frac{d_{L}d_{LS}}{d_{S}}
\nonumber
\\
&\simeq& 3\times(1+z_{L})\biggl(\frac{\sigma_{v}}{\rm 1m/s}\biggr)^{4}
\biggl(\frac{\hbar\omega}{\rm 1 keV}\biggr)\biggl(\frac{d_{L}d_{LS}/d_{S}}
{\rm 1Mpc}\biggr)
\nonumber
\\
&\simeq& 0.01(1+z_{L})\biggl(\frac{\sigma_{v}}{1{\rm km/s}}
\biggr)^{4}
\biggl(\frac{\nu}{\rm 1 Hz}\biggr)\biggl(\frac{d_{L}d_{LS}/d_{S}}
{1{\rm Gpc}}\biggr).
\label{MNEQN26}
\end{eqnarray}

For the SIS lens model, from Eq.~(\ref{MNEQN12}), the amplification 
factor is written analytically (see Appendix~A for derivation),
\begin{eqnarray}
F(w,y)=
e^{\frac{i}{2}w 
y^{2}}\sum_{n=0}^{\infty}\frac{\Gamma(1+\frac{n}{2})}{n!}
\left(2we^{i{3}\pi/2}\right)^{{n}/{2}}
{}_{1}F_{1}\biggl(1+\frac{n}{2},\hspace{1mm}1;\hspace{1mm}
-\frac{i}{2}wy^{2}\biggr).\label{MNEQN24}
\end{eqnarray}
Hence, the magnification is written as 
\begin{eqnarray}
\mu(w,y)=
\Big|\sum_{n=0}^{\infty}\frac{\Gamma(1+\frac{n}{2})}{n!}
\left(2we^{i{3}\pi/2}\right)^{{n}/{2}}
{}_{1}F_{1}\biggl(-\frac{n}{2},\hspace{1mm}1;\hspace{1mm}\frac{i}{2}
wy^{2}\biggr)\Big|^{2}.\label{MNEQN27}
\end{eqnarray}
The maximum magnification is given by setting $y=0$, 
\begin{eqnarray}
\mu_{max}&=&
\Big|\sum_{n=0}^{\infty}\frac{\Gamma(1+\frac{n}{2})}{n!}
\left(2we^{i{3}\pi/2}\right)^{{n}/{2}}\Big|^{2}
\label{MNEQN28}\\
&=&\Big|1+\frac{1}{2}(1-i)e^{-\frac{i}{2}w}\sqrt{\pi w}
\biggl[1+{\rm Erf}\biggl(\frac{\sqrt{w}}{2}
(1-i)\biggr)\biggr]\Big|^{2},\label{MNEQN29}
\end{eqnarray}
where ${\rm Erf}(z)$ is the error function (see Appendix A). 

Figure \ref{fig:four} shows the magnification $\mu(w,y)$ for the SIS model 
as a function of $w$ with the dimensionless source position fixed as 
$y=0.1,~0.5$, and $1$, respectively. 
Note that when $y=1$ (a single image is formed in the geometrical
optics limit), the oscillatory behavior appears.
Our result is consistent with the previous result 
\cite{Thesis,TNSIS}. 

\begin{figure}[t]
\epsfbox{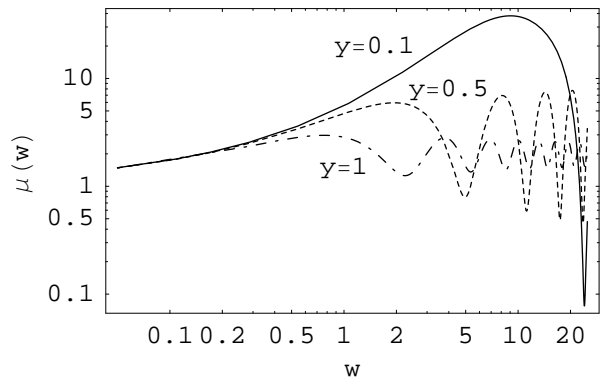}
\caption{\label{fig:four}
Same as figure \ref{fig:two} but for the SIS lens model. 
The behavior is very similar to that in Figure \ref{fig:two}.
}
\vspace{0.2cm}
\end{figure}

\begin{figure}
\epsfbox{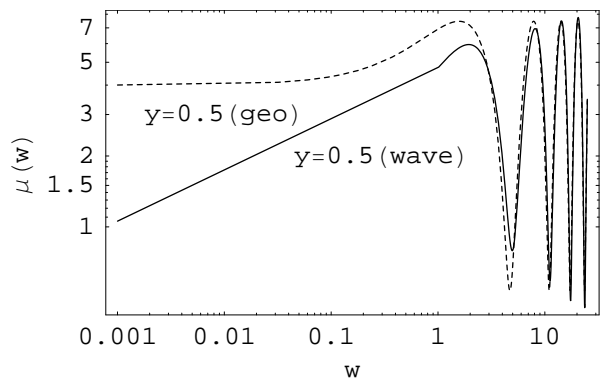}
\caption{\label{fig:five}
Same as figure \ref{fig:three} but for the SIS lens model.
The solid curve is the magnification $\mu(w,y)$ and 
the dashed curve is the corresponding geometrical optics 
formula $\mu_{geo}(w,y)$. Here, the source position is fixed 
as $y=0.5$. The geometrical optics approximation is not valid
for $w \simlt 1$.}
\vspace{0.2cm}
\end{figure}

Finally in this section let us consider the amplification factor
based on the geometrical optics estimation. 
In the SIS lens model, the two stationary 
points (the minimum and the saddle points) appear for $y<1$, while only one 
stationary point appears for  $y \geq 1$. Therefore we have
\begin{eqnarray}
F_{geo}(w,y)=
\left\{\begin{array}{rrr} 
|\mu_{+}|^{1/2}e^{(-iw(y+1/2))}
-i |\mu_{-}|^{1/2}e^{(iw(y+1/2))}
\hspace{4mm}(y<1),&&\\
&&\\
|\mu_{+}|^{1/2} \hspace{6.15cm}(y\geq 1),\label{MNEQN30}&&
\end{array}\right.,
\end{eqnarray}
from Eq.~(\ref{MNEQN15}), where $\mu_{\pm}=\pm 1+1/y$.
Then, the magnification in the geometrical optics is written 
\begin{eqnarray}
\mu_{geo}(w,y)=
\left\{\begin{array}{rrr} 
  {2}/{y}+2\sqrt{-1+1/y^{2}}\hspace{1mm}
\sin(2wy) \hspace{4mm}(y<1),&&\\
&&\\
1+{1}/{y} \hspace{3.87cm}(y \geq 1).\label{MNEQN31}
\end{array}\right.
\end{eqnarray}
Figure \ref{fig:five} compares the magnification (\ref{MNEQN27}) 
and (\ref{MNEQN31}) as a function of the parameter $w$, 
where the source position is fixed as $y=0.5$.

\section{Magnification of an Extended Source}
In this section we investigate the finite source size effect
in the wave optics. We consider the magnification from an
extended source with a Gaussian distribution of the surface 
brightness. 
The analytic formulas for the magnification in the previous
section are useful in the investigation in this section.

\subsection{Formulation}
\begin{figure}[t]
\epsfbox{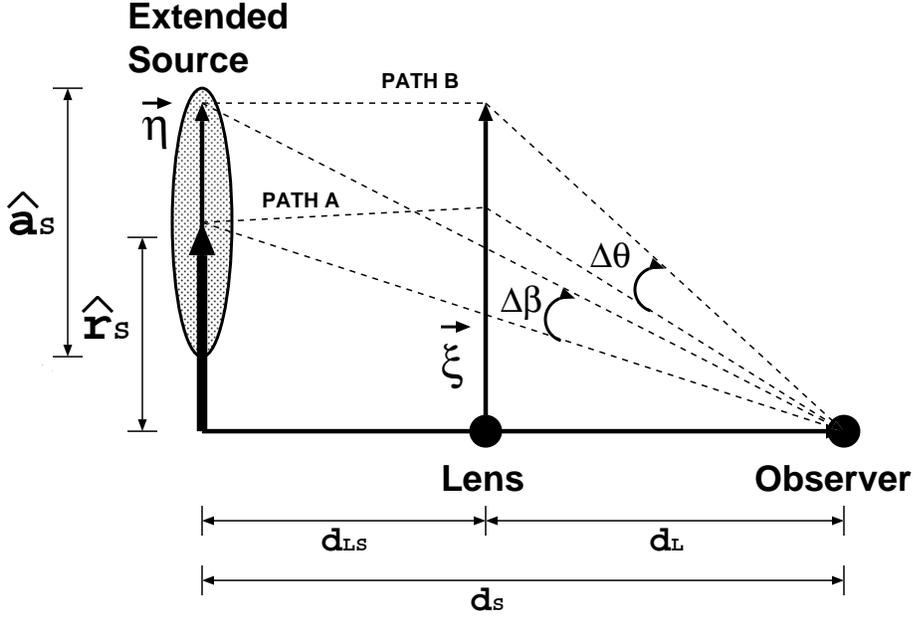}
\caption{\label{fig:six} 
Configuration of the gravitational lens system for an extended source. 
Here, $d_{L},d_{S}$ and $d_{LS}$ are the angular diameter 
distances between the observer and the lens, between the 
observer and the source, between the lens and the sources,
respectively. 
$\vec{\xi}$ and $\vec{\eta}$ are the dimensional
coordinates in the lens and the source planes, respectively. 
$\hat{r}_{S}$ specifies the position of the 
source center, and $\hat{a}_{S}$ is the source size. 
This sketch is based on the thin lens approximation.} 
\vspace{0.2cm}
\end{figure}

Following Ref. \cite{Krzysztof}, we consider the 
integral of the point source magnification weighted by 
the source intensity 
\begin{eqnarray}
\bar{\mu}(w,a_{S},r_{S})
=\frac{\int_{-\infty}^{\infty}W({\vec{y}})\mu(w,y)d^{2}y}
{\int_{-\infty}^{\infty}W({\vec{y}})d^{2}y},
\label{MNEQN32}
\end{eqnarray}
where we assume the Gaussian distribution of the source intensity 
\begin{eqnarray}
W({\vec{y}})
=\exp\biggl(-\frac{|\vec{y}-\vec{Y}|^{2}}{2a_{S}^{2}}\biggr),
\label{MNEQN33}
\end{eqnarray}
where $\vec{Y}$ $(|\vec{Y}|=r_{S})$ specifies the dimensionless 
source position, and $a_{S}$ is the dimensionless source size. 
These dimensionless quantities of the source position and the
source size are related to the dimensional quantities 
by
\begin{eqnarray}
r_{S}=\frac{\hat{r}_{S}}{d_{S}\theta_{E}}, 
\hspace{3mm}a_{S}=\frac{\hat{a}_{S}}{d_{S}\theta_{E}},\label{MNEQN34}
\end{eqnarray}
where $\hat{r}_{S}$ and $\hat{a}_{S}$ are the source position 
and the source size, respectively 
(see Figure \ref{fig:six}).
Note that the modified magnification depends on
the source size $a_{S}$ as well as the source position $r_S$.

We find the magnification can be written as
\begin{eqnarray}
\bar{\mu}(w,a_{S},r_{S})
=\frac{1}{a_{S}^{2}}e^{-\frac{r_{S}^{2}}{2a_{S}^{2}}}\int_{0}^{\infty}dy
\hspace{1mm}y\hspace{1mm}e^{-\frac{y^2}{2a_{S}^{2}}}I_{0}
\biggl(\frac{r_{S}}{a_{S}^{2}}y\biggr)\mu(w,y),\label{MNEQN35}
\end{eqnarray}
where $I_{0}(z)$ is the modified Bessel function of the zeroth order.

We will find the analytic expression for the integral (\ref{MNEQN35})
with the use of the result obtained in the previous section. 
Using the Taylor expansion of the magnification $\mu(w,y)$ around 
$y=r_{S}$,
\begin{eqnarray}
\mu(w,y)=\sum_{n=0}^{\infty}\frac{1}{n!}\mu^{(n)}(w,y=r_{S})(y-r_{S})^{n},
\label{MNEQN36}
\end{eqnarray}
where $\mu^{(n)}(w,y)$ is the n-rank derivative of $\mu(w,y)$ with respect 
to $y$,  the magnification (\ref{MNEQN35}) can be written in the form
\begin{eqnarray}
\bar{\mu}(w,a_{S},r_{S})
  &=&\sum_{n=0}^{\infty}A_{n}\hspace{1mm}\mu^{(n)}(w,y=r_{S}),
\label{MNEQN37}
\end{eqnarray}
where the coefficient is 
\begin{eqnarray}
A_{n}=
\frac{1}{a_{S}^{2}n!}e^{-\frac{r_{S}^{2}}{2a_{S}^{2}}}\int_{0}^{\infty}dy
\hspace{1mm}y\hspace{1mm}e^{-\frac{y^2}{2a_{S}^{2}}}
I_{0}\biggl(\frac{r_{S}}{a_{S}^{2}}y\biggr)(y-r_{S})^{n}.\label{MNEQN38}
\end{eqnarray}
We can evaluate the coefficients in an analytic manner. 
For example, for the first 
two terms, we have
\begin{eqnarray}
A_{0}&=&1,\label{MNEQN39}\\
A_{1}&=&a_{S}\biggl[-\frac{r_{S}}{a_{S}}
+\sqrt{\frac{\pi}{2}}e^{-\frac{r_{S}^{2}}{4a_{S}^{2}}}
\biggl(\biggl(1+\frac{r_{S}^{2}}{2a_{S}^{2}}\biggr)
I_{0}\biggl(\frac{r_{S}^{2}}{4a_{S}^{2}}
\biggr)+\frac{r_{S}^{2}}{2a_{S}^{2}}I_{1}
\biggl(\frac{r_{S}^{2}}{4a_{S}^{2}}
\biggr)\biggr)\biggr]\label{MNEQN40}
\\  
&\simeq&\left\{
\begin{array}{cc}
a_S^2/r_S   &(r_S\gg a_S)\\
a_S \sqrt{\pi/2}  &(r_S\ll a_S)
\end{array}
\right.,
\end{eqnarray}
where $I_{1}(z)$ is the modified Bessel function of the first order. 
The other terms can be evaluated in a similar way.
Note that the zeroth order term of Eq.~(\ref{MNEQN37}) 
reproduces the magnification of the point source. Hence, our 
expression (\ref{MNEQN37}) is based on the expansion around 
the point source limit. 

\begin{figure}[t]
\epsfbox{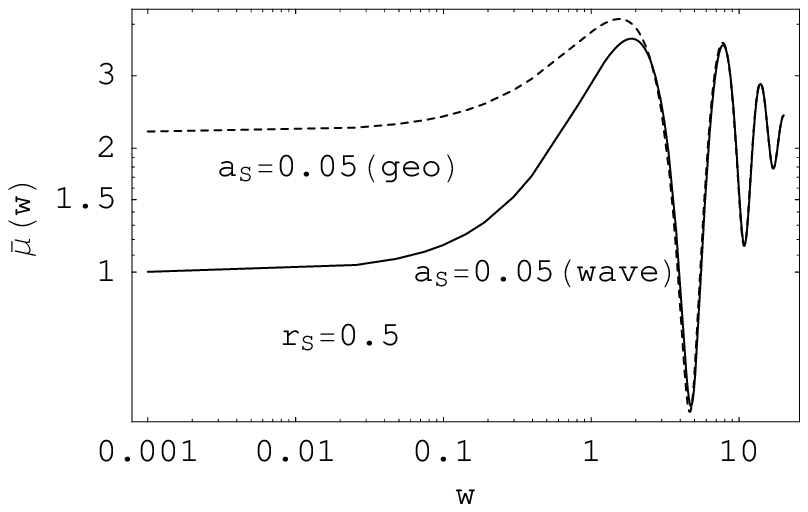}
\caption{\label{fig:seven}
The magnification of the extended source 
$\bar{\mu}(w,a_{S},r_{S})$ for the point 
mass lens model, as a function $w$.
Here, the position of the source center and the source radius 
are fixed $r_{S}=0.5$ and $a_{S}=0.5$.
For $w\simgt 1$, the wave optics agrees with the geometrical optics.}
\vspace{0.5cm}
\epsfbox{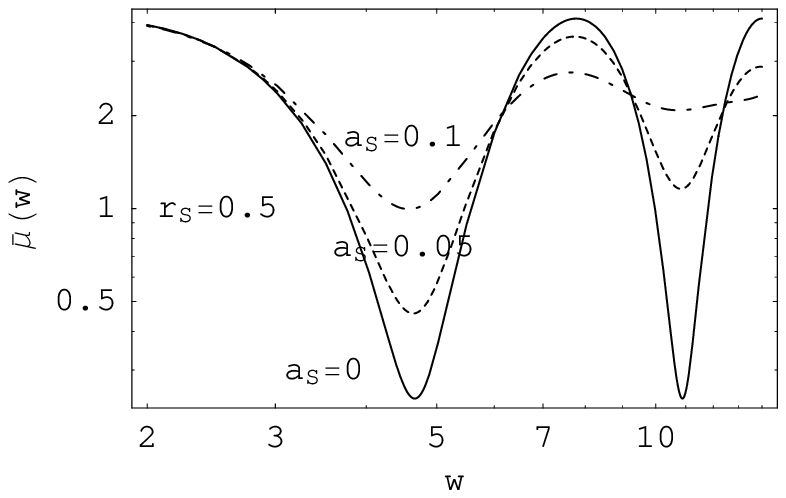}
\caption{\label{fig:eight}
The magnification $\bar{\mu}(w,a_{S},r_{S})$ as a function of $w$
for the point mass lens model. Here, the position of the 
source center is fixed as $r_{S}=0.5$, and the three curves
show the different source sizes specified by $a_{S}=0,~~a_{S}=0.05$ 
and $a_{S}=0.1$, respectively. As the source size becomes larger, 
the oscillation-amplitude of the magnification decreases.
In the computation of this magnification, we used 
the approximation of the geometrical optics. This provides
a good approximation as long as $w \simgt 1$, as 
demonstrated in figure \ref{fig:seven}.
}
\vspace{0.2cm}
\end{figure}

\subsection{Point Mass Lens}
For the point mass lens case, with the use of the 
expression (\ref{MNEQN20}), we can evaluate 
the magnification of the extended source.
Here, we write the first two terms,
\begin{eqnarray}
\bar{\mu}(w,a_{S},r_{S})&\simeq&
A_{0}\mu^{(0)}(w,r_{S})+A_{1}\mu^{(1)}(w,r_{S})
\label{MNEQN41}\\
&=&\mu(w,r_{S})\Biggl[1-A_{1}w^2r_{S}
\Re\Biggl[\frac{{}_{1}F_{1}(1+\frac{i}{2}w,\hspace{1mm}2;\hspace{1mm}
\frac{i}{2}w r_{S}^{2})}
{{}_{1}F_{1}(\frac{i}{2}w,\hspace{1mm}1;\hspace{1mm} 
\frac{i}{2}w r_{S}^{2})}\Biggr]\Biggr].\label{MNEQN42}
\end{eqnarray}

For the point mass lens model, with the use of the 
approximate expression (\ref{MNEQN23}), the magnification of 
the extended source is evaluated in the geometrical optics. 
Figure \ref{fig:seven} shows $\bar{\mu}(w,a_{S},r_{S})$ and
the corresponding magnification with the geometrical optics approximation.
Here, in evaluating $\bar{\mu}(w,a_{S},r_{S})$, we summed the terms 
up to $n=20$, and the position of the source center and the radius of the source are 
fixed as $a_{S}=0.5$ and $r_{S}=0.5$, respectively. Note that both the
curves agree for $w\simgt 1$.  Comparing it with Figure \ref{fig:three}, 
the oscillation-amplitude decreases as $w$ becomes large.

Figure \ref{fig:eight} plots the magnification of the extended source,
as a function of $w$. Here, the source position is fixed as 
$r_{S}=0.5$, and the three curves assume the source size
$a_{S}=0$, $a_{S}=0.05$, and $a_{S}=0.1$, respectively. 
As the source size becomes larger, the oscillation-amplitude 
of the magnification decreases. 
This can be understood as follows. The oscillation feature comes from 
the interference of two waves in the geometrical optics in the case of the
point source. In the case of an extended source, the wave magnification 
is determined by a superposition of many waves.
Then, the clear interference disappears by averaging over the phase.

Now let us evaluate the condition that the finite source size
effect becomes substantial in an analytic manner. The ratio 
of the second term to the first term of the right hand side 
of Eq.~(\ref{MNEQN42}) is
\begin{eqnarray}
\delta \bar{\mu} / {\mu} 
&=&-A_{1}w^2 r_{S}
\Re\Biggl[\frac{{}_{1}F_{1}(1+\frac{i}{2}w,\hspace{1mm}2;\hspace{1mm}
\frac{i}{2}w r_{S}^{2})}
{{}_{1}F_{1}(\frac{i}{2}w,\hspace{1mm}1;\hspace{1mm} \frac{i}{2}w 
r_{S}^{2})}\Biggr]
\label{MNEQN43}\\
&\simeq& -w^{2}r_{S}A_{1}+{\cal O}(w^4).
\label{MNEQN44}
\end{eqnarray} 
The condition that the 
finite source size effect becomes substantial is 
$|\delta \bar \mu/\mu |\sim {\cal O} (1)$.
For the case $r_{S}\gg a_{S}$, we may approximate 
$A_{1}\simeq a_{S}^{2}/r_{S}$, and we have
\begin{eqnarray}
\delta \bar{\mu}/{\mu}\simeq -(a_{S}w)^{2}.
\label{delmumua}
\end{eqnarray}
Thus, $a_{S}w$ is a key parameter of the finite source size effect
in the wave optics.

Next, let us examine the finite source size effect near the 
caustic $r_S=0$ in detail. Some aspects have been discussed in Ref. \cite{DN}. 
The above argument is based on the expansion of the magnification 
in terms of $w$, which is not suitable for the large value of 
$w$.  We here consider the regime of the geometrical optics. 
In the limit $w\gg1$ and $y\ll 1/w^{1/2}$, we may
write 
\begin{eqnarray}
  \mu(w,y)={\pi w} J_0(wy)^2,
\label{JJ}
\end{eqnarray}
where we used the mathematical formula \cite{Magnus}
\begin{eqnarray}
  \lim_{a\rightarrow \infty} {}_1F_1(a,1;z/a)= I_0(2\sqrt{z})
\end{eqnarray}
with $z$ fixed. Note that the approximate formula (\ref{JJ}) 
is rather general, which can be derived from (\ref{MNEQN12})
with the saddle point method for $w\gg1$ (\cite{DN}, see also 
below). 

Substituting (\ref{JJ}) into (\ref{MNEQN35}), 
the magnification can be evaluated as
\begin{eqnarray}
\bar{\mu}(w,a_{S},r_{S})
=\frac{\pi w}{a_{S}^{2}}e^{-\frac{r_{S}^{2}}{2a_{S}^{2}}}\int_{0}^{\infty}dy
\hspace{1mm}y\hspace{1mm}e^{-\frac{y^2}{2a_{S}^{2}}}I_{0}
\biggl(\frac{r_{S}}{a_{S}^{2}}y\biggr)J_0(wy)^2,
\label{jjj}
\end{eqnarray}
which is valid for $r_S<a_S\ll 1/w^{1/2}\ll1$.  Using the definition of
the modified Bessel function
\begin{eqnarray}
I_0(z)=\sum_{m=0}^\infty {z^{2m}\over (m!)^2 2^{2m}},
\end{eqnarray}
we can write
\begin{eqnarray}
\bar{\mu}(w,a_{S},r_{S})
=\frac{\pi w}{a_{S}^{2}}e^{-\frac{r_{S}^{2}}{2a_{S}^{2}}}
\sum_{m=0}^\infty (-1)^m{(r_S/a_S^2)^{2m}\over (m!)^2 2^{2m}}
{\partial^m  \alpha_0(\beta)\over \partial \beta^m}\bigg|_{\beta=1/2a_S^2},
\label{betai}
\end{eqnarray}
where we defined
\begin{eqnarray}
  \alpha_0(\beta)&=&\int_{0}^{\infty}dy 
\hspace{1mm}y\hspace{1mm}e^{-\beta{y^2}} J_0(wy)^2
={1\over 2\beta} e^{-w^2/2\beta} I_0(w^2/2\beta). 
\end{eqnarray}
Using the condition, $r_S<a_S\ll 1/{w}^{1/2}\ll 1$, 
the first two terms of (\ref{betai}) yield
\begin{eqnarray}
\bar{\mu}(w,a_{S},r_{S})
\simeq
 \pi w e^{-r_S^2/2a_S^2} e^{-w^2a_S^2} I_0(w^2a_S^2)
\left(1+{1\over 2}{r_S^2\over a_S^2}\right),
\end{eqnarray}
which reduces to 
\begin{eqnarray}
\bar{\mu}(w,a_{S},r_{S})
\simeq\left\{
\begin{array}{ll}
\displaystyle{
  \pi w e^{-r_S^2/2a_S^2} \left(1+{1\over 2}{r_S^2\over a_S^2}\right)}
\\
\vspace{3.5mm}
  \hspace{1cm}
  =\displaystyle{\pi w\left(1+{\cal O}(r_S^4/a_S^4)\right)}
  &~~~(wa_S\ll1)
\\
\displaystyle{\sqrt{\pi\over 2}{1\over a_S} e^{-r_S^2/2a_S^2} 
\left(1+{1\over 2}{r_S^2\over a_S^2}\right)}
\\
  \hspace{1cm}  =\displaystyle{\sqrt{\pi\over 2}{1\over a_S}
 \left(1+{\cal O}(r_S^4/a_S^4)\right)}
  &~~~(wa_S\gg1)
\end{array}
\right..
\label{jkj}
\end{eqnarray}
This expression means the followings: the result of the point source 
is reproduced for $wa_S\ll1$ and $r_S=0$, while 
the finite source size effect becomes substantial 
for $wa_S\gg1$ and $\bar \mu$ approaches to the 
constant value $\sqrt{\pi/2}/a_S$ for $r_S=0$. 
The magnification $\bar \mu$ becomes smaller as 
$r_S/a_S$ becomes large.

\begin{figure}[t]
\epsfbox{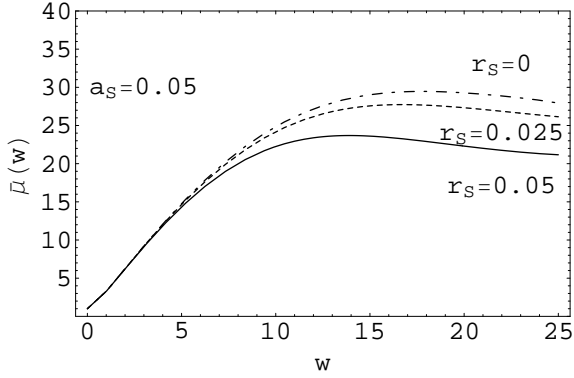}
\caption{\label{fig:nine} 
Magnification $\bar{\mu}(w,a_{S},r_{S})$ as a function of $w$. 
Here, the source size is fixed as $a_{S}=0.05$, and 
the position of the source center is $r_{S}=0,~0.025$, 
and $0.05$, respectively.}
\vspace{0.2cm}
\end{figure}
\begin{figure}
\epsfbox{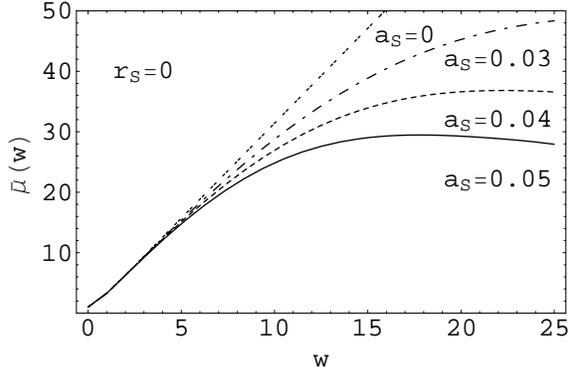}
\caption{\label{fig:extrafig2} 
Magnification $\bar{\mu}(w,a_{S},r_{S})$ as a function of $w$. 
Here, the position of the source center is fixed as 
$r_{S}=0$, and the source size is $a_{S}=0,~0.03,~0.04$, 
and $0.05$, respectively.}
\vspace{0.2cm}
\end{figure}

Figure \ref{fig:nine} shows $\bar{\mu}(w,a_{S},r_{S})$ 
as a function of $w$ with the source size $a_{S}=0.05$ 
for the different source position 
$r_{S}=0,~0.025$, and $0.05$, respectively. 
In this figure we see that $\bar{\mu}$ approaches a constant value
$(\sim 1/a_S)$ slightly depending on $r_S$,   
as $w$ becomes large.  
Figure \ref{fig:extrafig2} plots $\bar{\mu}(w,a_{S},r_{S})$ 
with the source position fixed as $r_{S}=0$, and 
the different source size $a_{S}=0,~0.03,~0.04$, and 
$0.05$, respectively. 
Even for the Einstein ring configuration, 
due to the finite source size
effect,  the maximum amplification is limited by the 
the factor $(\sim 1/a_S)$ for $w \simgt 1/a_S$.
These behaviors are consistent with those expected from the 
above analytic argument.

The above results demonstrate that the condition 
that the finite source size effect becomes 
substantial is $wa_S>1$. The reason can be
understood as follows:
As shown in Appendix B, the condition, $wa_S=1$, is 
equivalent to the condition that the path difference 
between the PATH A and the PATH B in Figure \ref{fig:six} 
becomes comparable to the wavelength. 
Therefore, the observed wave is a superposition 
of many waves with different phases for $wa_S\gg1$. 
This eliminates the interference feature and decreases 
the oscillation feature in the energy spectrum. 
The finite source size effect near the caustic 
$r_S=0$ is also understood in the similar way. 
The maximum magnification is decreased by 
averaging over the magnification of different
phases that depend on the position on the source surface.

\subsection{Singular Isothermal Sphere Lens}
\begin{figure}[t]
\epsfbox{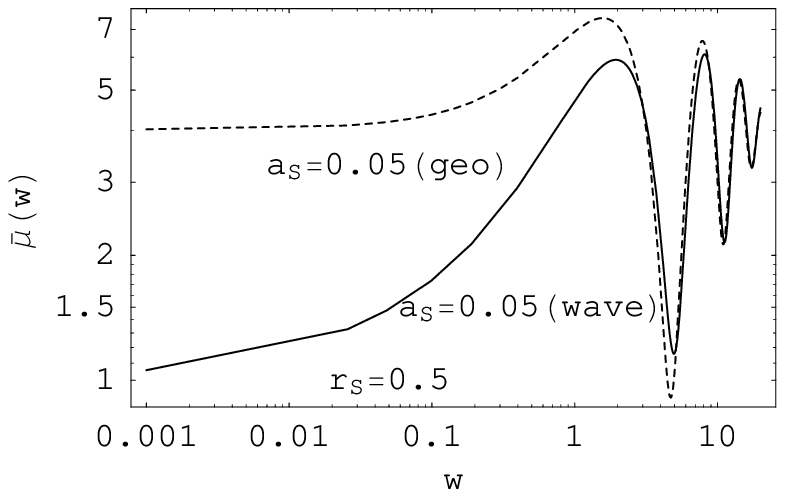}
\caption{\label{fig:eleven}
Same as Figure \ref{fig:seven}, but for the SIS lens model.
For $w\simgt 5$, the difference between the wave optics
and the geometrical optics is negligible.
Here, we fixed $r_{S}=0.5$ and $a_{S}=0.5$.}
\vspace{0.2cm}
%
\epsfbox{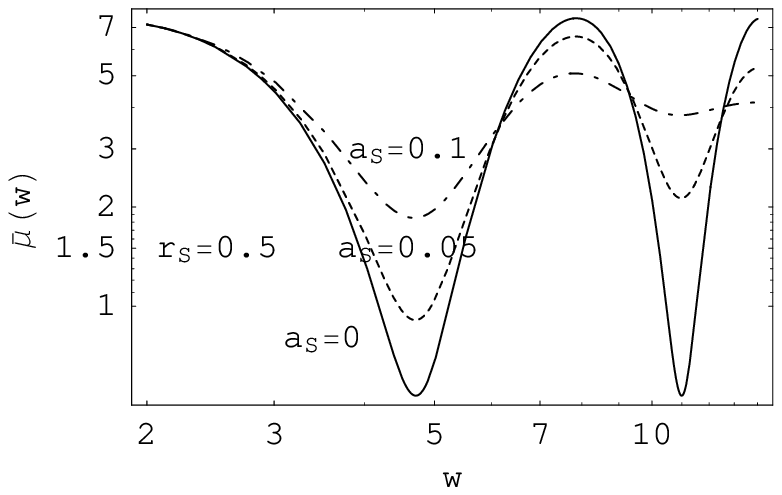}
\caption{\label{fig:twelve}
Same as Figure \ref{fig:eight}, but for the SIS lens model.
Here, the position of the source center is fixed $r_S=0.5$,
and the magnification is plotted for $a_S=0,~0.05$, and $0.1$,
respectively.
}
\vspace{0.2cm}
\end{figure}

We here consider the finite 
source size effect in the SIS lens model. In this case, the 
magnification of the extended source (\ref{MNEQN37}) can be 
evaluated with the expression (\ref{MNEQN27}). 
Figure \ref{fig:eleven} shows the magnification $\bar \mu$ 
(solid curve), where we set $r_S=0.5$ and $a_S=0.5$. 
The dashed curve is the corresponding geometrical optics 
with (\ref{MNEQN31}) instead of (\ref{MNEQN27}). 
In the numerical computation, we performed the sum with respect 
to $n$ up to $20$. 

Figure \ref{fig:twelve} plots the magnification of the extended 
source, as a function of $w$. Here, the source position is fixed as 
$r_{S}=0.5$, and the three curves assume the source size
$a_{S}=0$, $a_{S}=0.05$, and $a_{S}=0.1$, respectively. 
As the source size becomes larger, the oscillation-amplitude 
of the magnification decreases. Similarly to the case of the point mass lens
model,
the clear oscillation feature disappears as the source size becomes large.
In this figure we used the approximation of the geometrical optics 
in evaluating the magnification because of a convenience
of numerical technique. The validity of its approximation is 
demonstrated in Figure \ref{fig:eleven} at lease 
for $w \simgt 5$. 
The approximation is not very good for $w \simlt 5$, 
however, it does not alter our conclusions.

We write down the first two terms of the magnification
\begin{eqnarray}
\bar{\mu}(w,r_{S},a_{S})
&\simeq& 
A_{0}\mu^{(0)}(w,r_{S})+A_{1}\mu^{(1)}(w,r_{S}),\label{MNEQN46}\\
&=&\mu(w,r_{S})\Biggl[1-A_{1}w r_{S}
\nonumber\\
&\times&
\Re\Biggl[\frac{\sum_{n=0}^{\infty}
\frac{\Gamma(1+\frac{n}{2})}{n!}(-i n)
 \left(2we^{i{3}\pi/2}\right)^{n/2}
{}_{1}F_{1}(1-\frac{n}{2},\hspace{1mm}2;\hspace{1mm}\frac{i}{2}w 
r_{S}^{2})}
{\sum_{n=0}^{\infty}\frac{\Gamma(1+\frac{n}{2})}{n!}
 \left(2we^{i{3}\pi/2}\right)^{n/2}
{}_{1}F_{1}(-\frac{n}{2},
\hspace{1mm}1;\hspace{1mm} 
\frac{i}{2}w r_{S}^{2})}\Biggr]\Biggr].
\nonumber\\
\label{MNEQN47}
\end{eqnarray}
The ratio of the second term to the first term of the 
right hand side of Eq.~(\ref{MNEQN47}) is 
\begin{eqnarray}
\delta \bar{\mu}/{\mu}&=&-A_{1}w r_{S}
\nonumber\\
&&\times
\Re\Biggl[\frac{\sum_{n=0}^{\infty}
\frac{\Gamma(1+\frac{n}{2})}{n!}(-i n)
\left(2we^{i{3}\pi/2}\right)^{n/2}
 {}_{1}F_{1}(1-\frac{n}{2},\hspace{1mm}2;\hspace{1mm}\frac{i}{2}w 
r_{S}^{2})}
{\sum_{n=0}^{\infty}\frac{\Gamma(1+\frac{n}{2})}{n!}
\left(2we^{i{3}\pi/2}\right)^{n/2}
{}_{1}F_{1}(-\frac{n}{2},\hspace{1mm}1;\hspace{1mm} \frac{i}{2}w 
r_{S}^{2})}\Biggr]
\label{MNEQN48}
\\
&\simeq& -{\sqrt{\pi}\over 2}w^{3/2}r_{S}A_{1}
+\left(2-{\pi\over 2}\right)w^2 r_S A_1  +{\cal O}(w^{5/2}).
\label{MNEQN49}
\end{eqnarray}
The condition that the point source approximation breaks is 
$|\delta \bar{\mu}/ {\mu}|\sim {\cal O}(1)$.
In the limit $r_{S}\gg a_{S}$, we have $A_{1}\simeq a_{S}^{2}/r_{S}$,
then 
\begin{eqnarray}
\delta \bar{\mu}/{\mu}\simeq -{\sqrt{\pi}\over 2}a_{S}^2w^{3/2}
+\left(2-{\pi\over 2}\right)w^2 a_s^2  +{\cal O}(w^{5/2}). 
\label{delmumub}
\end{eqnarray}

\begin{figure}[t]
\epsfbox{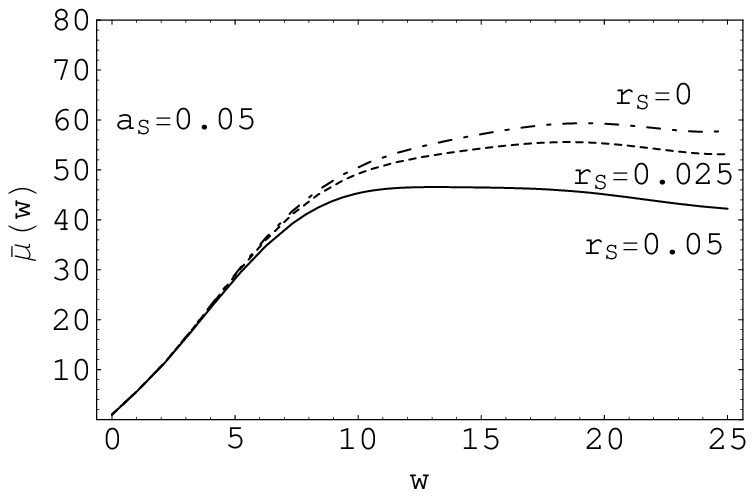}
\caption{\label{fig:thirteen}
Same as Figure \ref{fig:nine}, but for the SIS lens model.} 
\vspace{0.2cm}
\end{figure}

\begin{figure}[t]
\epsfbox{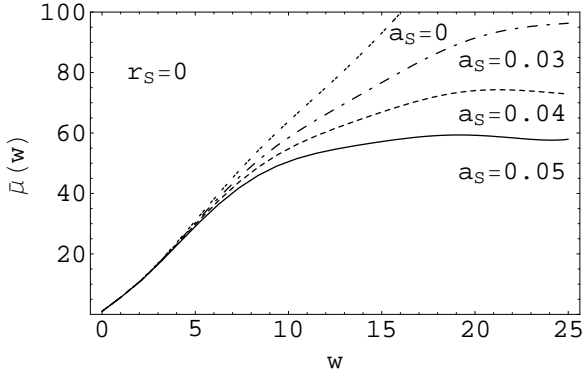}
\caption{\label{fig:extrafig4}
Same as Figure \ref{fig:extrafig2}, but for the SIS lens model.} 
\vspace{0.2cm}
\end{figure}

Next, similarly to the point mass lens model, we
examine the finite source size effect near the 
caustic $r_S=0$. Here, let us consider the
approximate estimation of Eq.~(\ref{MNEQN12})
using the saddle point method, as demonstrated 
in Ref. \cite{DN}. Using the approximate method,
for $w\gg 1$ and $y\ll 1/\sqrt{w}$, we obtain 
\begin{eqnarray}
  \mu(w,y)\simeq {{2\pi w}x_*^2 \over |{1-\psi''(x_*)}|}J_0(wx_*y)^2,
\end{eqnarray}
where $x_*$ is a positive solution of the lens equation 
$x=\psi'(x)$. 
For the SIS model, we have  
\begin{eqnarray}
\mu(w,y)\simeq2\pi wJ_0(wy)^2. 
\label{joj}
\end{eqnarray}
Substituting Eq.~(\ref{joj}) into (\ref{MNEQN35}), we have
the same expressions of the magnification as Eqs. (\ref{jjj})
and (\ref{jkj}), but with multiplied by the constant factor $2$.

Figure \ref{fig:thirteen} shows $\bar{\mu}(w,a_{S},r_{S})$ 
as a function of $w$ with the source size $a_{S}=0.05$ 
for the different source position 
$r_{S}=0,~0.025$, and $0.05$, respectively, for the SIS lens model.
Similarly, Figure \ref{fig:extrafig4} plots $\bar{\mu}(w,a_{S},r_{S})$ 
with the source position fixed as $r_{S}=0$, and the 
different source size $a_{S}=0,~0.03,~0.04$, and 
$0.05$, respectively. 
These figures show the similar behaviors to those in 
the point mass lens model.

\begin{figure}[t]
\epsfbox{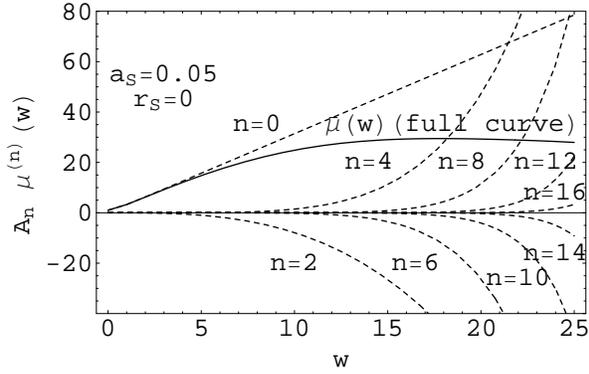}
\caption{\label{fig:seriesfig1}
 $A_n\mu^{(n)}$ in Eq.~(\ref{MNEQN37}) as a function of $w$
 for the point mass lens model.
 The dashed curves are the terms for $n=0$, $2$, $4$, $6$, $8$, $12$, $14$, and 
 $16$, respectively, while the solid curve is the sum up to the term for $n=20$. 
 Here, we fixed the parameters as $a_S=0.05$ and $r_S=0$. 
}
\vspace{0.2cm}
\end{figure}

\begin{figure}[t]
\epsfbox{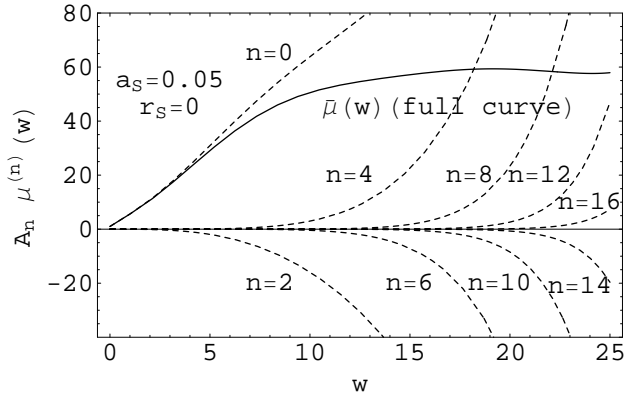}
\caption{\label{fig:seriesfig2}
Same as Figure \ref{fig:seriesfig1}, but for the SIS lens model.} 
\vspace{0.2cm}
\end{figure}

Finally in this section, we mention our computation 
and the numerical convergence. We have performed 
the numerical computation using the package MATHEMATICA.
The terms in Eq.~(\ref{MNEQN37}) with respect to $n$ 
are summed up to $n=20$. The dashed curves in 
Figure \ref{fig:seriesfig1} shows each term 
of $A_n\mu^{(n)}$ as a function $w$ for the point mass lens model, 
where we adopted the parameters $a_S=0.05$ and $r_S=0$. 
The solid curve is the (summed) magnification $\bar \mu(w)$. 
Figure \ref{fig:seriesfig2} is same as Figure \ref{fig:seriesfig1} but for the
SIS lens model. As long as $w\simlt20$, the convergence 
of our computation is evident. But for the large value of 
$w$, our method is not advantageous because higher terms
with respect to $n$ is required. The numerical methods
developed in Refs. \cite{DN,Thesis,TNSIS} would be 
useful for numerical computation of general cases.

\section{Discussion}
In this section, let us discuss astrophysical consequences 
of the result in the previous section. 
We first summarize the conditions that the lensing 
signature of the wave optics may appear in the 
spectral feature, as follows:
\begin{eqnarray}
w \sim 1,
\label{cond1}
\\
|\delta \bar\mu/\mu| 
\simlt 1.
\label{cond2}
\end{eqnarray} 
The first condition (\ref{cond1}) is that the wave optics in 
lensing becomes important for a
monochromatic wave from a point 
source. The second condition (\ref{cond2}) is that the 
oscillation feature in energy spectra survives 
against the finite source size effect. Here, we consider the 
case $r_S\simgt a_S$. Therefore, with Eqs.~(\ref{delmumua})
and (\ref{delmumub}), combination of both the condition 
simply gives $a_S\simlt 1$.

Then, we discuss possible observational consequences of the wave 
effect in the astrophysical situation, the lensing of the 
gravitational wave from a binary compact objects \cite{DN} and
the {\it femtolensing} of the gamma ray burst \cite{Krzysztof,Gould}. 

\subsection{Gravitational Wave from a Compact Binary}
For the point mass lens model, the dimensionless parameter $w$ 
is given by Eq.~(\ref{MNEQN19}).
The condition $w\sim 1$ yields 
\begin{eqnarray}
\biggl(\frac{\nu}{\rm 1Hz}\biggr)
\sim {0.8\over 1+z_L}
\biggl(\frac{M}{10^4 M_{\odot}}\biggr)^{-1}.
\label{MNEQNA}
\end{eqnarray}
We also have 
\begin{eqnarray}
a_S\simeq 1\times 10^{-11}
\biggl({\hat a_S\over 10^3{\rm km}}\biggr)
\biggl(\frac{M}{10^4 M_{\odot}}\biggr)^{-1/2}
\biggl({{\rm H_{0}^{-1}}\over d_{LS}d_{S}/d_L }\biggr)^{1/2},
\label{MNEQNB}
\end{eqnarray}
where ${\rm H_{0}^{-1}}=(70{\rm km/s/Mpc})^{-1}
=1.3 \times 10^{26}{\rm m}$ is the Hubble distance.
On the other hand, for the SIS model, 
$w$ is given by (\ref{MNEQN26}). Then, from 
$w\sim1$, we have
\begin{eqnarray}
\biggl(\frac{\nu}{\rm 1Hz}\biggr)
\sim {20\over 1+z_L}
\biggl(\frac{\sigma_v}{1{\rm km/s}}\biggr)^{-4}
\left({d_L d_{LS}/d_S\over {\rm H_{0}^{-1}}}\right)^{-1}.
\label{MNEQN52}
\end{eqnarray}
We also have 
\begin{eqnarray}
a_S\simeq 6\times 10^{-11}
\biggl({\hat a_S\over 10^3{\rm km}}\biggr)
\biggl({\sigma_v\over1{\rm km/s}}\biggr)^{-2}
\biggl({{\rm H_{0}^{-1}}\over d_{LS}}\biggr).
\label{MNEQNH}
\end{eqnarray}
Let us consider the binary compact objects, which consists 
of the two equal objects with the mass $m$. For the 
binary we have the relation
$(G m)^{2}/2L=4^{-2/3}(G m)^{5/3}(\omega/2)^{2/3}$, where 
$L$ is the distance between the two objects and $\omega$
is the angular frequency of the gravitational wave. 
This equation is rewritten as
\begin{eqnarray}
L\simeq 3\times10^{3}\biggl(\frac{m}{M_{\odot}}\biggr)^{1/3}
\biggl(\frac{\nu}{\rm 1Hz}\biggr)^{-2/3}\hspace{2mm}{\rm km}.
\label{MNEQN53}
\end{eqnarray}
If we set $L\sim \hat a_S$,
it is clear that $a_S\ll 1$
as long as we consider the gravitational wave of the 
frequency around $1 {\rm Hz}$.
This means that the point source approximation is 
very good for the gravitational wave of this frequency.
\subsection{Femtolensing}
We next consider the {femtolensing}, which was 
pointed out by Gould, and Stanek, Paczynski and Goodman \cite{Krzysztof,Gould}.
The femtolensing is the lens effect by a tiny mass 
on the gamma ray burst.

For the point mass lens model, the condition $w\sim 1$ is
rewritten as 
\begin{eqnarray}
\biggl(\frac{h\nu}{\rm 1keV}\biggr) \sim {0.7\over 1+z_L}
\biggl(\frac{M}{10^{20}{\rm g}}\biggr)^{-1},
\label{MNEQND}
\end{eqnarray}
where $h$ is the Planck constant. 
The dimensionless source size may be written as 
\begin{eqnarray}
a_S\simeq 0.5\times 
\biggl({\hat a_S\over 10^5{\rm km}}\biggr)
\biggl(\frac{M}{10^{20}{\rm g}}\biggr)^{-1/2}
\biggl({ {\rm H_{0}^{-1}}\over d_{LS}d_{S}/d_L}\biggr)^{1/2}.
\label{MNEQNE}
\end{eqnarray}
This suggests that 
the finite source size effect is important
in the femtolensing by the point mass lens. If the
source size is larger than $10^5$km, the finite 
source size effect becomes significant. In this case
the signature of the interference in energy 
spectra will disappear. 

On the other hand, for the SIS model, 
(\ref{MNEQN52}) is rewritten as 
\begin{eqnarray}
\biggl(\frac{h\nu}{\rm 1keV}\biggr)
\sim {0.8\over 1+z_L}
\biggl(\frac{\sigma_v}{0.1{\rm m/s}}\biggr)^{-4}
\left({d_L d_{LS}/d_S\over {\rm H_{0}^{-1}}}\right)^{-1}.
\label{MNEQNG}
\end{eqnarray}
We also have 
\begin{eqnarray}
a_S\simeq 0.5\times
\biggl({\hat a_S\over 10^5{\rm km}}\biggr)
\biggl({\sigma_v\over 0.1{\rm m/s}}\biggr)^{-2}
\biggl({{\rm H_{0}^{-1}}\over d_{LS}}\biggr).
\label{MNEQNF}
\end{eqnarray}
This suggests that the femtolensing might have occurred 
due to a very small mass halo, if it existed. 
Such the very small mass halo might be unrealistic 
in our universe, (cf. \cite{Kolb}). 
However, Moore et al. have pointed 
out the possibility of the survival of very small 
mass halos which are produced in the high redshift 
universe, depending on the dark matter model 
\cite{MooreA}, though the possibility is still open 
to debate \cite{MooreB,ZhaoA,ZhaoB}. Our investigation
suggests that the source size is a crucial factor 
even if the femtolensing occurred by such the very small halo. 
When the source size is larger than $10^5$km, the interference 
signature will be significantly affected by the finite source 
size effect.

\subsection{Finite Source Size Effect near the Caustic}
We now consider the finite source size effect in 
the wave optics near the caustic. We have demonstrated 
that is becomes influential when $a_S w\simgt 1$ for 
$a_S\ll 1/\sqrt{w}\ll 1$. For the gravitational wave from 
a compact binary, $a_S w\ll1$ will be reasonable for general 
situation.
Hence, we here consider the femtolensing. We may write 
\begin{eqnarray}
  a_Sw\simeq 0.7\times(1+z_L) 
  \biggl(\frac{h\nu}{\rm 1keV}\biggr)
  \biggl(\frac{M}{10^{20}{\rm g}}\biggr)^{1/2}
  \biggl({\hat a_S\over 10^5{\rm km}}\biggr)
  \biggl({ {\rm H_{0}^{-1}}\over d_{LS}d_{S}/d_L}\biggr)^{1/2},
\label{addfla}
\end{eqnarray}
for the point mass lens mode, 
\begin{eqnarray}
a_Sw\simeq 0.7\times(1+z_L) 
  \biggl(\frac{h\nu}{\rm 1keV}\biggr)
 \biggl({\sigma_v\over 0.1{\rm m/s}}\biggr)^2
\biggl({\hat a_S\over 10^5{\rm km}}\biggr)
\biggl({d_L\over d_{S}}\biggr),
\label{addflb}
\end{eqnarray}
for the SIS lens model, respectively. These estimations
suggest that the finite source size effect is important
in the femtolensing near the caustic too. 

\section{Summary and Conclusions}
In this paper we investigated the finite source size 
effect on the wave optics in the gravitational lensing. 
First we presented the analytic expression of the 
magnification for the SIS lens model as well as the point 
mass lens model. Based on the result, we evaluated 
the magnification of the finite-size source, assuming a 
Gaussian profile for the surface intensity.
The analytic expression of the magnification is given 
in terms of the expansion with respect to the source size. 
This expression is useful to understand how the finite 
source size effect works on the spectral feature of
the magnification in the wave optics.
The condition that the finite source size effect becomes
significant is discussed. 
As application of the result, we considered the 
finite source size effect on the wave optics in lensing
of the gravitational wave from a compact binary and the
femtolensing. For the lensing of the gravitational wave, 
it is demonstrated that the finite source size effect 
can be negligible as long as we consider the gravitational wave 
of the frequency around $1$Hz.
For the femtolensing of the gamma ray burst, we 
confirmed the result by Stanek et al. \cite{Krzysztof}, 
for the point mass lens model.
We also considered the femtolensing by the hypothetically 
very small halo. The femtolensing might imprint the
lensing signature on energy spectra if occurred, however,
the finite source size effect is crucial. If the source
size is larger than $10^5$km, the finite source size effect 
becomes significant and will not allow the detection of 
the interference signature in the energy spectra. 

It is worthy noting the finite source size effect 
near the caustic. In the wave optics of the lens configuration of 
the Einstein ring, the maximum magnification of the point source 
is not divergent, but is in proportion to the frequency of the wave. 
But, for an extended sources, the finite source size effect 
becomes substantial for 
$a_Sw
\simgt1$, 
in which the maximum magnification is suppressed by the value
$\sqrt{\pi/2}/a_S$ as long as $a_S\ll 1/\sqrt{w}\ll1$. 
This finite source size effect would be influential in 
the femtolensing near the caustic too,
if the source size is larger than $10^5$km.

\ack
All the numerical computation presented in this paper were performed with 
the help of the package MATHEMATICA version 5.0. 
The authors thank an anonymous referee for useful comments
which helped improve the manuscript.
We are also grateful to Y. Kojima, R. Yamazaki, M. Sakagami, K. Nakao, 
C. Yoo and R. Takahashi for useful comments and conversations 
related to the topic in the present paper.

\section*{References}

\vspace{1cm}
\appendix
\section{Analytic Expression for the Amplification Factor in the SIS Model}
In this appendix, we derive an analytic expression of the 
amplification factor of the SIS lens model. 
Since the gravitational deflection potential is $\psi(x)=x$, 
Eq.~(\ref{MNEQN12}) is written
\begin{eqnarray}
F(w,y)&=&-iw e^{{i}wy^2/2}\int_{0}^{\infty}dx
\hspace{1mm}x\hspace{1mm}
J_{0}(wxy)\exp\Biggl[iw\Biggl(\frac{1}{2}x^{2}-x \Biggr)\Biggr] 
\label{MNEQN A.1}\\
&=&-iw e^{{i}w y^2/2}\sum_{n=0}^{\infty}\frac{(-iw)^{n}}{n!}
\int_{0}^{\infty}dx\hspace{1mm}x^{1+n}\hspace{1mm}
J_{0}(wxy)e^{{i}w x^{2}/2}. \label{MNEQN A.2}
\end{eqnarray}
Using a mathematical formula \cite{Abramowitz}, 
Eq.~(\ref{MNEQN A.2}) is integrated as
\begin{eqnarray}
F(w,y)&=&e^{{i}w y^{2}/2}\sum_{n=0}^{\infty}
\frac{\Gamma(1+\frac{n}{2})}{n!}
\left(2we^{{i3}\pi/2}\right)^{{n}/{2}}
{}_{1}F_{1}
\biggl(1+\frac{n}{2},\hspace{1mm}1;\hspace{1mm}-\frac{i}{2}w y^{2}\biggr),
\label{MNEQN A.3}
\end{eqnarray}
where ${}_1F_1(a,c,z)$ is the confluent hypergeometric function 
\cite{Magnus}. Using the formula 
\begin{eqnarray}
e^{-z}{}_{1}F_{1}\biggl(-\frac{n}{2},\hspace{1mm}1;\hspace{1mm}z\biggr)
={}_{1}F_{1}\biggl(1+\frac{n}{2},\hspace{1mm}1;\hspace{1mm}-z \biggr), 
\label{MNEQN A.4}
\end{eqnarray}
we have
\begin{eqnarray}
F(w,y)=\sum_{n=0}^{\infty}\frac{\Gamma(1+\frac{n}{2})}{n!}
\left(2we^{i3\pi/2}\right)^{{n}/{2}}
{}_{1}F_{1}\biggl(-\frac{n}{2},\hspace{1mm}1;\hspace{1mm}
\frac{i}{2}w y^{2}\biggr).\label{MNEQN A.5}
\end{eqnarray}

In the case $y=0$, the Einstein ring configuration,
the amplification factor is 
\begin{eqnarray}
F(w,y=0)&=&-iw\int_{0}^{\infty}dx\hspace{1mm}x\hspace{1mm}
\exp \biggl(iw\frac{x^{2}}{2}-iw x \biggr) \label{MNEQN A.6}
\\
&=&w\int_{0}^{\infty}dt\hspace{1mm}t\hspace{1mm}
\exp \biggl(-\frac{w}{2}t^{2}-i^{{3}/{2}}wt \biggr).
\hspace{5mm}\label{MNEQN A.7}
\end{eqnarray} 
Using a mathematical formula \cite{Gradshteyn}, we have the
analytic simple form 
\begin{eqnarray}
F(w,y=0)=e^{-{i}w/4}D_{-2}
\biggl(e^{i{3}\pi/4}\sqrt{w}\biggr),\label{MNEQN A.8}
\end{eqnarray}
where $D_{-2}(z)$ is the parabolic cylinder function. 
With the use of the error function, defined by 
\begin{eqnarray}
{\rm Erf}(z)=\frac{2}{\sqrt{\pi}}\int_{0}^{z}dt\hspace{1mm}e^{-t^{2}},
\end{eqnarray} 
we may write
\begin{eqnarray}
D_{-2}(z)=\sqrt{\frac{\pi}{2}}e^{{z^{2}}/{4}}z
\biggl(1-{\rm Erf} \biggl( \frac{z}{\sqrt{2}} \biggr)\biggr)
-e^{-{z^{2}}/{4}},\label{MNEQN A.9} 
\end{eqnarray}
and we finally have
\begin{eqnarray}
F(w,y=0)=1+\frac{1}{2}(1-i)e^{-{i}w/2}\sqrt{\pi w}
\biggl[1+{\rm Erf}\biggl(\frac{\sqrt{w}}{2}
(1-i)\biggr)\biggr].\label{MNEQN A.10}
\end{eqnarray}

\section{Path Difference and the Finite Source Size Effect}
Here, we consider the path difference between the PATH A
and PATH B in Figure \ref{fig:six}, and show that
$wa_S=1$ is equivalent to the condition that the path 
difference is comparable to the wavelength.
Using the angle of the unlensed source position $\vec{\beta}$ 
and the angle of the image position $\vec{\theta}$, the 
Fermat's potential is given 
\begin{eqnarray}
\hat{\phi}(\theta,\beta)=\frac{d_{L}d_{S}}{2d_{LS}}(\vec{\theta}
-\vec{\beta})^{2}
-\hat{\psi}(\theta).\label{MNEQN B.1}
\end{eqnarray}
In the case $\vec \beta =(\beta,0)$ and $\vec\theta=(\theta,0)$,
the path difference can be evaluated by 
\begin{eqnarray}
\Delta\hat{\phi}(\theta,\beta)
=\frac{d_{L}d_{S}}{d_{LS}}(\theta-\beta)(\Delta\theta-\Delta\beta)
-\hat\psi'(\theta)\Delta\theta,
\label{MNEQN B.2}
\end{eqnarray}
where we assume $\Delta \theta \ll \theta$ and $\Delta \beta\ll \beta$,
and $\hat\psi'(\theta)={d\hat{\psi}(\theta)}/{d\theta}$, (see Figure \ref{fig:six}
for the definitions of $\Delta \theta$ and $\Delta \beta$).

The gravitational lens equation is given by 
\begin{eqnarray}
\frac{d_{L}d_{S}}{d_{LS}}(\theta-\beta)
-{\hat{\psi}'(\theta)}=0,
\label{MNEQN B.22}
\end{eqnarray}
and the Einstein angle is defined by the solution of the equation
\begin{eqnarray}
\frac{d_{L}d_{S}}{d_{LS}}\theta_E
-{\hat{\psi}'(\theta_E)}=0.
\label{MNEQN B.3}
\end{eqnarray}
With the use of the above equations, we have
\begin{eqnarray}
\Delta\hat{\phi}(\theta,\beta)
=-\hat\psi'(\theta)\Delta\beta.
\label{MNEQN B.23}
\end{eqnarray}
We adopt the approximation for the phase difference,
\begin{eqnarray}
\Delta\hat{\phi}(\theta,\beta)
&\simeq&-\hat\psi'(\theta_E)\Delta\beta
\simeq-\frac{d_{L}d_{S}}{d_{LS}}\theta_E\Delta\beta.
\label{MNEQN B.24}
\end{eqnarray}
Because $\Delta \beta=\hat a_S/d_L$, the condition 
$|(1+z_L)\Delta\hat \phi|=\lambda/2\pi$ yields 
\begin{eqnarray}
\omega(1+z_L)\frac{d_{L}}{d_{LS}} \hat a_S\theta_E(=a_S w)=1,
\label{MNEQN B.25}
\end{eqnarray}
where $\lambda$ is the wavelength of the propagating wave. 
\end{document}